\documentclass[twoside,12pt]{article}
\usepackage{epsfig}

\def\Journal#1#2#3#4{{#1} {#2} (#4) #3 }

\def\NPA{{\em Nucl. Phys.} A}

\def\PLB{{\em Phys. Lett.} B}

\def\PRL{\em Phys. Rev. Lett.}
\def\PREV{\em Phys. Rev.}
\def\PREP{\em Phys. Rep.}

\def\PRD{{\em Phys. Rev.} D}
\def\PRC{{\em Phys. Rev.} C}

\def\EPJC{{\em Eur. Phys. J.} C}
\def\JPG{{\em J. Phys.} G}

\newcommand{\be}{\begin{equation}}
\newcommand{\ee}{\end{equation}}
\newcommand{\bea}{\begin{eqnarray}}
\newcommand{\eea}{\end{eqnarray}}

\begin{document}

\title{ \vspace{1cm} Thermal dileptons in high-energy nuclear collisions}
\author{Sanja Damjanovic\\
CERN, 1211 Geneva 23, Switzerland}

\maketitle
\begin{abstract}
Clear signs of excess dileptons above the known sources were found at
the SPS since long. However, a real clarification of these
observations was only recently achieved by NA60, measuring dimuons
with unprecedented precision in 158A GeV In-In collisions. The excess
mass spectrum in the region $M$$<$1 GeV is consistent with a dominant
contribution from $\pi^{+}\pi^{-} \rightarrow \rho \rightarrow
\mu^{+}\mu^{-}$ annihilation. The associated $\rho$ spectral function
shows a strong broadening, but essentially no shift in mass. In the
region $M$$>$1 GeV, the excess is found to be prompt, not due to
enhanced charm production. The inverse slope parameter
$T_\mathrm{eff}$ associated with the transverse momentum spectra rises
with mass up to the $\rho$, followed by a sudden decline above. While
the initial rise, coupled to a hierarchy in hadron freeze-out, points
to radial flow of a hadronic decay source, the decline above signals a
transition to a low-flow source, presumably of partonic origin. The
mass spectra show the steep rise towards low masses characteristic for
Planck-like radiation. The polarization of the excess referred to the
Collins Soper frame is found to be isotropic. All observations are
consistent with a global interpretation of the excess as thermal
radiation. We conclude with a short discussion of a possible link to
direct photons.
\end{abstract}

\vspace*{0.4cm}
\noindent{\bf Introduction}
\vspace*{0.2cm}

Dileptons are particularly attractive to study the hot and dense QCD
matter formed in high-energy nuclear collisions. In contrast to
hadrons, they directly probe the entire space-time evolution of the
expanding system, escaping freely without final-state interactions. At
low masses $M$$<$1 GeV (LMR), thermal dilepton production is mediated
by the broad vector meson $\rho$ (770) in the hadronic phase. Due to
its strong coupling to the $\pi\pi$ channel and the short life time of
only 1.3 fm/c, ``in-medium'' modifications of its mass and width close
to the QCD phase boundary have since long been considered as the prime
signature for {\it chiral symmetry
restoration}~\cite{Pisarski:mq,Rapp:1995zy,Brown:kk}. At intermediate
masses $M$$>$1 GeV (IMR), it has been controversial up to today
whether thermal dileptons are dominantly produced in the earlier
partonic or in the hadronic phase, based here on hadronic processes
other than $\pi\pi$ annihilation. Originally, thermal emission from
the early phase was considered as a prime probe of {\it
deconfinement}~\cite{McLerran:1984ay,Kajantie:1986}.

Excess dileptons above the known decay sources at SPS energies were
observed before by CERES \cite{Agakichiev:1995xb,CERES:2008} for
$M$$<$1 GeV, NA38/NA50~\cite{Abreu:2002rm} for $M$$>$1 GeV and by
HELIOS-3~\cite{Masera:1995ck} for both mass regions
(see~\cite{Specht:2007ez} for a short recent review including the
preceding $pp$ era and the theoretical milestones). The sole existence
of an excess gave a strong boost to theory, with hundreds of
publications. In the LMR region, $\pi\pi$ annihilation with
regeneration and strong in-medium modifications of the intermediate
$\rho$ during the fireball expansion emerged as the dominant
source. However, the data quality in terms of statistics and mass
resolution remained largely insufficient for a precise assessment for
the in-medium spectral properties of the $\rho$. In the IMR region,
thermal sources or enhanced charm production could account for the
excess equally well, but that ambiguity could not be resolved, nor
could the nature of the thermal sources be clarified.

\begin{figure}[]
\begin{center}
\includegraphics*[width=0.43\textwidth]{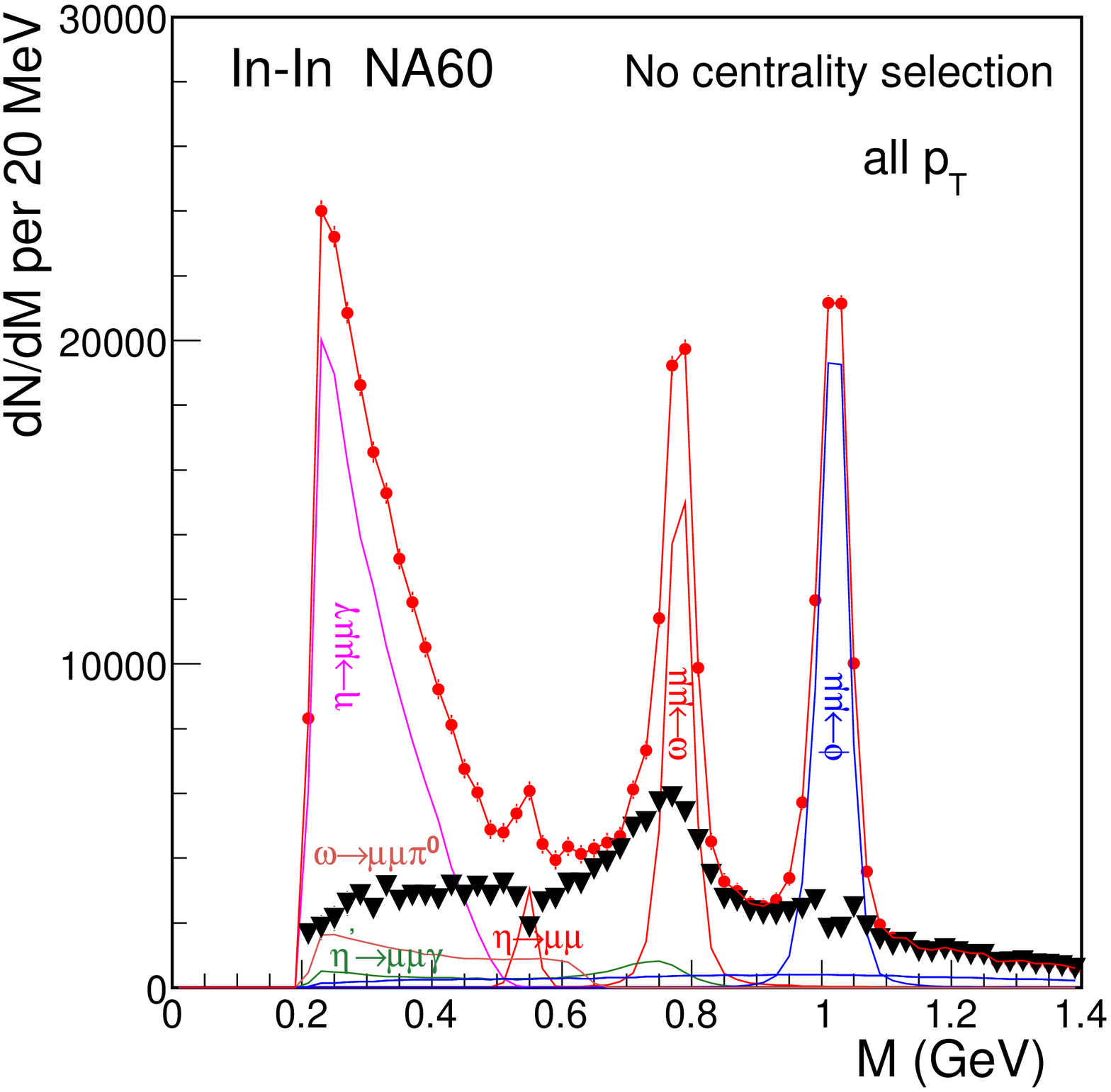}
\hspace*{0.2cm}
\includegraphics*[width=0.43\textwidth]{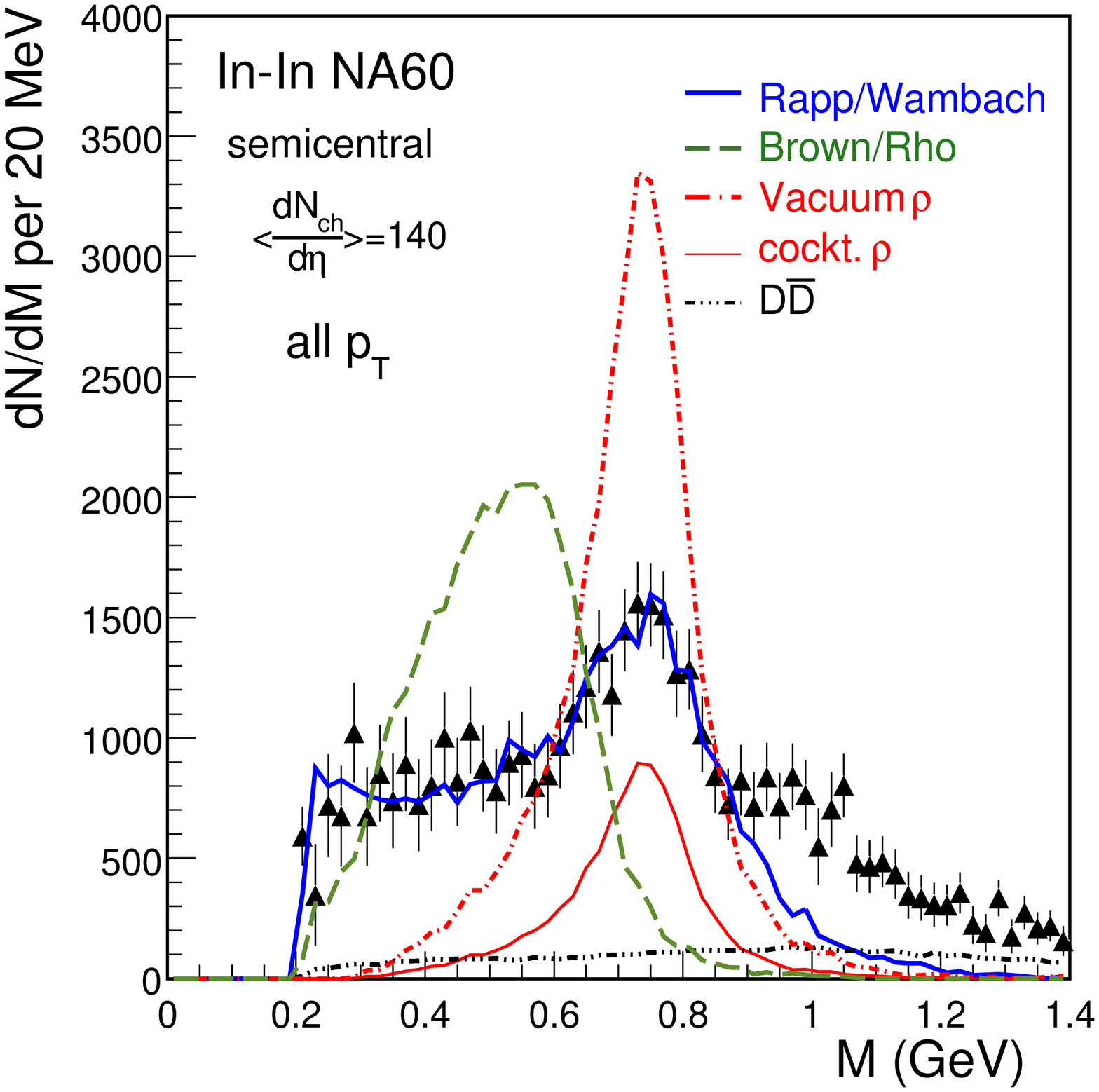}
\caption{Left: Background-subtracted mass spectrum before (dots) and
after subtraction of the known decay sources (triangles). Right:
Excess dimuons compared to theoretical predictions~\cite{Rapp:2005pr},
renormalized to the data in the mass interval M$<$0.9 GeV. No
acceptance correction applied.}
   \label{fig1}
\end{center}
\end{figure}

A big step forward in technology, leading to completely new standards
of the data quality in this field, has recently been achieved by NA60,
a third-generation experiment built specifically to follow up the open
issues addressed above~\cite{NA60:2008IMR}. Results on mass and
transverse momentum spectra of excess dimuons have already been
published~\cite{NA60:2008IMR,Arnaldi:2006jq,Arnaldi:2007ru}. Further
aspects associated with the centrality dependence, polarization and
acceptance-corrected mass spectra were reported during
2008~\cite{Damjanovic:qm08,Damjanovic:hp08}. This paper shortly
reviews the central results and also discusses a possible link of the
excess dileptons at very low masses to real photons.

\vspace*{0.4cm}
\noindent{\bf Mass and p$_{T}$ spectra of excess dileptons }
\vspace*{0.2cm}

Fig.~\ref{fig1} (left) shows the centrality-integrated net dimuon mass
spectrum for 158A GeV In-In collisions in the LMR region. The narrow
vector mesons $\omega$ and $\phi$ are completely resolved; the mass
resolution at the $\omega$ is 20 MeV. The peripheral data can fully be
described by the electromagnetic decays of neutral
mesons~\cite{Arnaldi:2006jq,Damjanovic:2006bd}. This is not true for
the more central data as plotted in Fig.~\ref{fig1}, due to the
existence of a strong excess of pairs. The high data quality of NA60
allows to isolate this excess with {\it a priori unknown
characteristics} without any fits: the cocktail of the decay sources
is subtracted from the total data using {\it local} criteria, which
are solely based on the mass distribution itself. The $\rho$ is not
subtracted. The excess resulting from this difference formation is
illustrated in the same figure
(see~\cite{Arnaldi:2006jq,Arnaldi:2007ru,Damjanovic:2006bd} for
details and error discussion).

The common features of the excess mass spectra can be recognized in
Fig.~\ref{fig1} (right). A peaked structure is always seen, residing
on a broad continuum with a yield strongly increasing with centrality
(see below), but remaining essentially centered around the nominal
$\rho$ pole~\cite{Damjanovic:2006bd}. Without any acceptance
correction and $p_{T}$ selection, the data can directly be interpreted
as the {\it space-time averaged spectral function} of the $\rho$, due
to a fortuitous cancellation of the mass and $p_{T}$ dependence of the
acceptance filtering and the phase space factors associated with
thermal dilepton emission~\cite{Damjanovic:2006bd}. The two main
theoretical scenarios for the in-medium spectral properties of the
$\rho$, broadening~\cite{Rapp:1995zy} and dropping
mass~\cite{Brown:kk}, are shown for comparison~\cite{Rapp:2005pr}.
Clearly, the broadening scenario gets close, while the dropping mass
scenario in the version which described the CERES data reasonably
well~\cite{Rapp:1995zy,Brown:kk,Agakichiev:1995xb} fails for the much
more precise NA60 data. 

\begin{figure}[h!]
\begin{minipage}[l][6.5cm][t]{8cm}
\includegraphics*[scale=0.45]{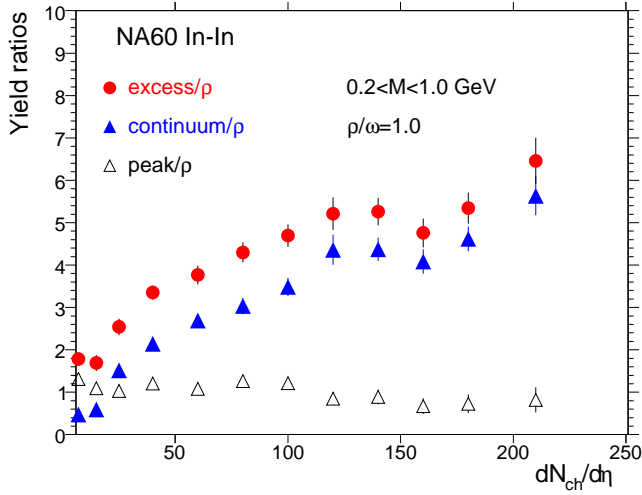}
\end{minipage} \qquad\qquad
\begin{minipage}[l][6.5cm][t]{8cm}
\vspace*{3.8cm}
\caption{Excess yield ratios for peak, continuum and total
vs. centrality for the mass window 0.2$<$$M$$<$1 GeV and all p$_{T}$.}
   \label{fig2}
\end{minipage}
\end{figure}

A detailed view of the shape of the excess mass spectra is obtained by
using a side window method~\cite{Damjanovic:2006bd} to determine
separately the yields of the peak and the underlying
continuum. Fig.~\ref{fig2} shows the centrality dependence of these
variables: peak, underlying continuum and total yield in the mass
interval 0.2$<$$M$$<$1.0 GeV, all normalized to the cocktail
$\rho$. The continuum and the total show a very strong increase,
starting already in the peripheral region, while the peak slowly
decreases from $>$1 to $<$1. Recalling that Fig.~\ref{fig1} roughly
represents the full $\rho$ spectral function, the excess/$\rho$ ratio
can directly be interpreted as the number of $\rho$ generations
created by formation and decay during the fireball evolution,
including freeze-out: the '$\rho$-clock', frequently discussed in the
past. It reaches up to about 6 generations for central In-In
collisions; selecting lower $p_{T}$ this number doubles.
\begin{figure}[h!]
\begin{center}
\includegraphics*[width=0.43\textwidth]{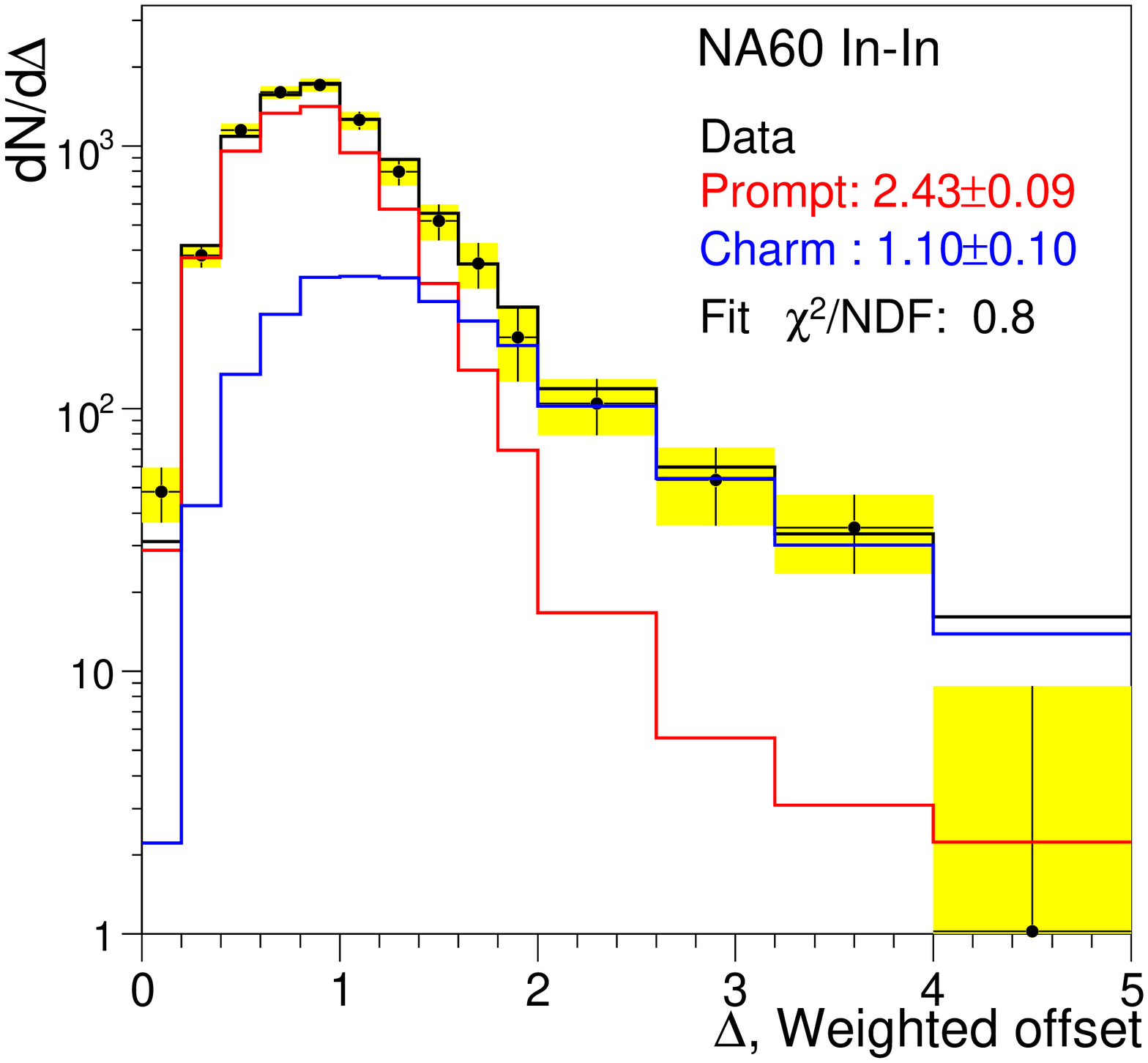}
\hspace*{0.7cm}
\includegraphics*[width=0.43\textwidth]{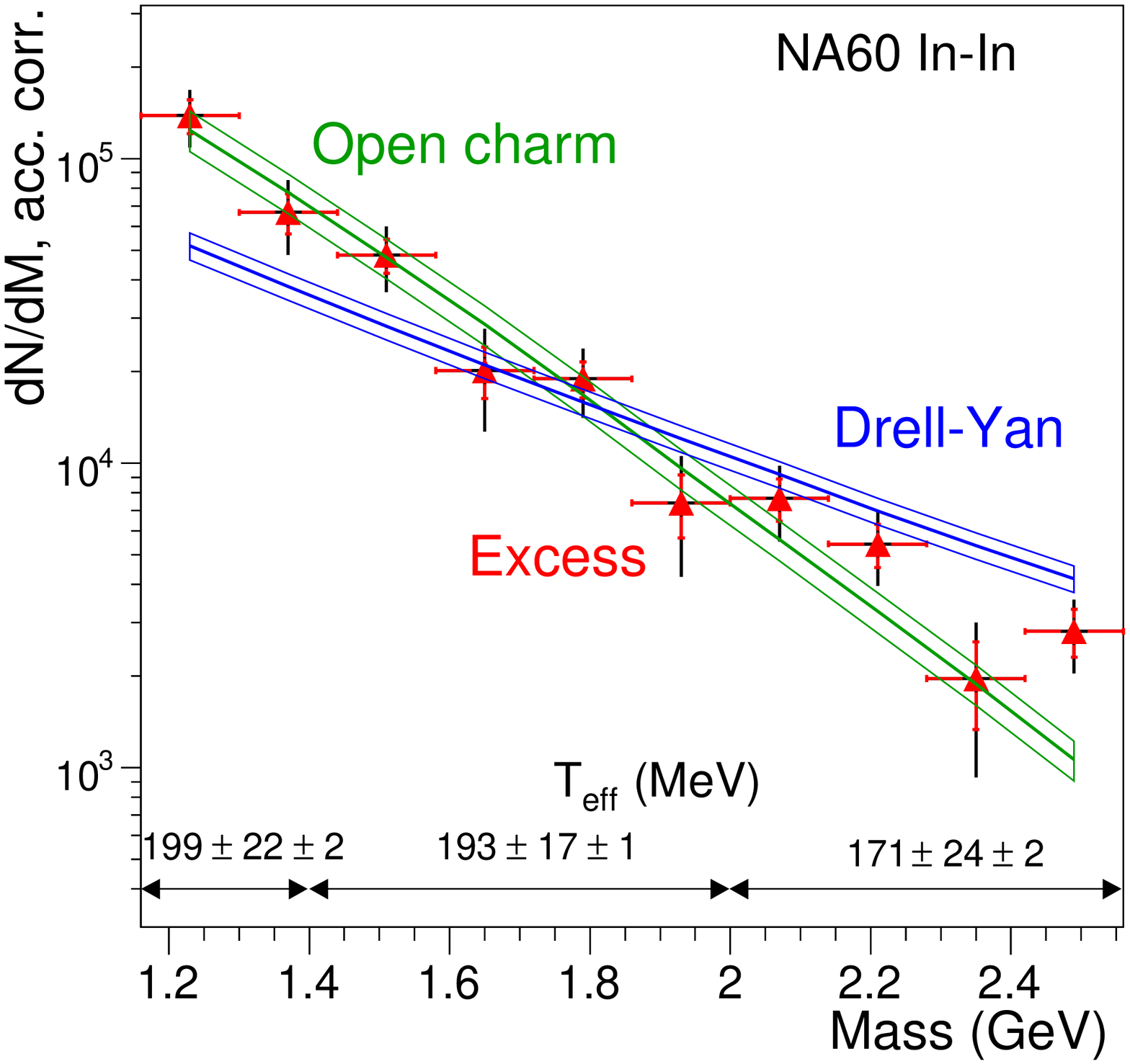}
\caption{Left: Fit of the weighted offset distribution in the IMR
region with the contributions from charm and prompt decays. Right:
Acceptance-corrected mass spectra of Drell-Yan, open charm and the
excess (triangles).}
   \label{fig3}
\end{center}
\end{figure}

The central NA60 results in the IMR region~\cite{NA60:2008IMR} are
shown in Fig.~\ref{fig3}. The use of the $Si$-vertex tracker allows to
measure the offset between the muon tracks and the main interaction
vertex and thereby to disentangle prompt and offset dimuons from $D$
decays. The left panel of Fig.~\ref{fig3} shows the offset
distribution to be perfectly consistent with no charm enhancement,
expressed by a fraction of $1.1\pm0.1$ of the canonical level. The
observed excess is really prompt, with an enhancement over Drell-Yan
by a factor of 2.4. The excess can now be isolated in the same way as
was done in the LMR region, subtracting the measured known sources,
here DY and open charm, from the total data. The right panel of
Fig.~\ref{fig3} shows the decomposition of the total into DY, open
charm and the prompt excess. The mass spectrum of the excess is quite
similar to the shape of open charm and much steeper than DY; this
explains of course why NA50 could describe the excess as enhanced open
charm. The transverse momentum spectra are also much steeper than
DY~\cite{NA60:2008IMR}. The fit temperatures of the $m_{T}$ spectra
associated with 3 mass windows are indicated on the bottom of the
figure.

The remainder of this paper is solely concerned with excess data fully
corrected for acceptance and pair
efficiencies~\cite{Arnaldi:2007ru,Damjanovic:2007qm}. In principle,
the correction requires a 4-dimensional grid in the space of
$M$-$p_{T}$-$y$-$cos{\theta_{CS}}$ (where $\theta_{CS}$ is the polar
angle of the muons in the Collins Soper frame). To avoid large
statistical errors in low-acceptance bins, it is performed instead in
2-dimensional $M$-$p_{T}$ space, using the measured $y$ and
$cos{\theta}$ distributions as an input. The latter are, in turn,
obtained with acceptance corrections determined in an iterative way
from MC simulations matched to the data in $M$ and $p_{T}$. The
$y$-distribution is found to have the same rapidity width as
$dN_{ch}/d\eta$, $\sigma_{y}\sim$ 1.5~\cite{Damjanovic:2007qm}. The
$cos{\theta_{CS}}$ distributions for two mass windows of the excess
and the $\omega$ are contained in~\cite{Damjanovic:qm08}. Within
errors, they are found to be uniform, implying the polarization of the
excess dimuons to be zero, in contrast to DY and consistent with the
expectations for thermal radiation from a randomized system.

\begin{figure*}[]
\begin{center}
\includegraphics*[width=0.43\textwidth]{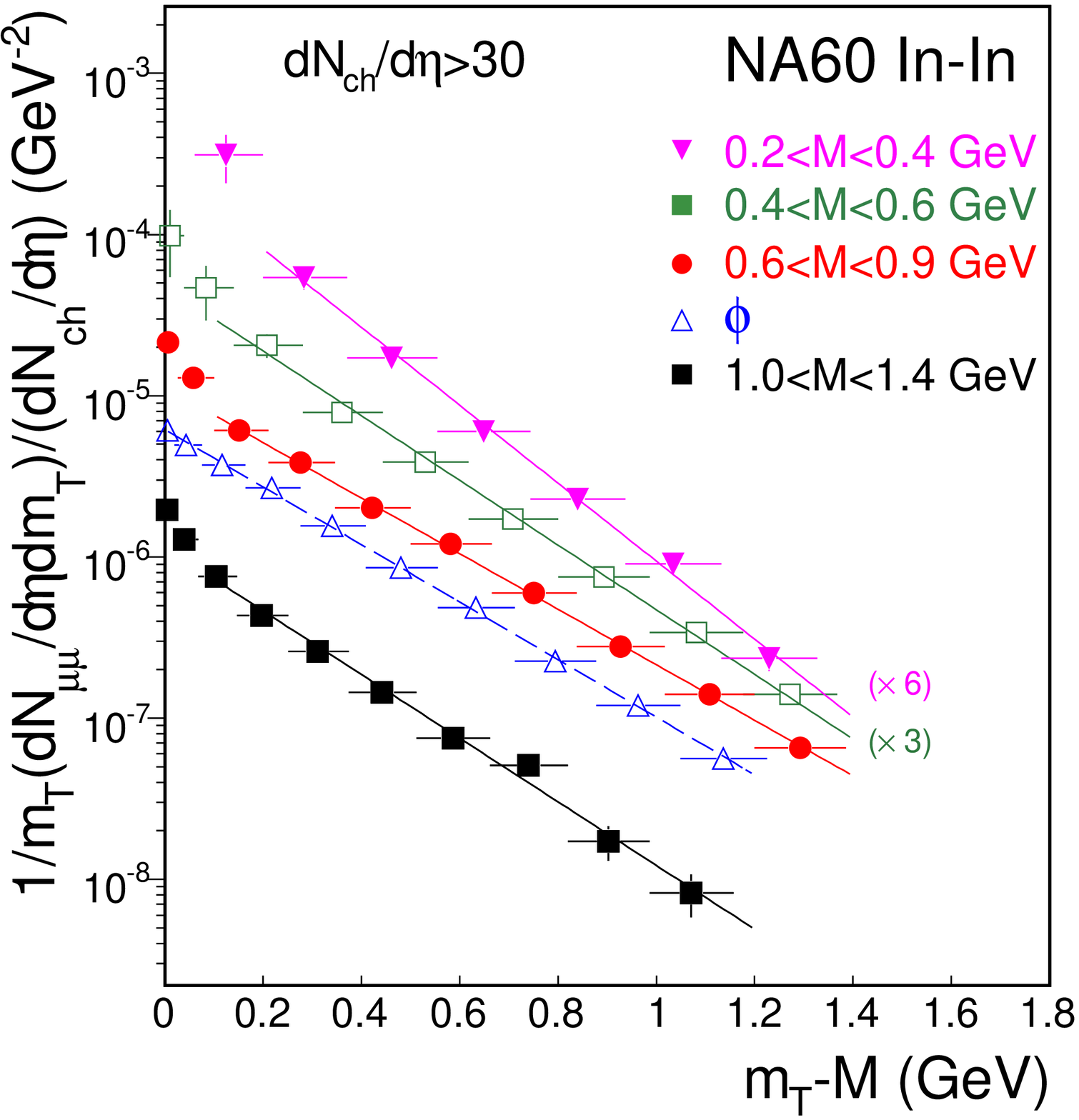}
\hspace*{0.7cm}
\includegraphics*[width=0.43\textwidth]{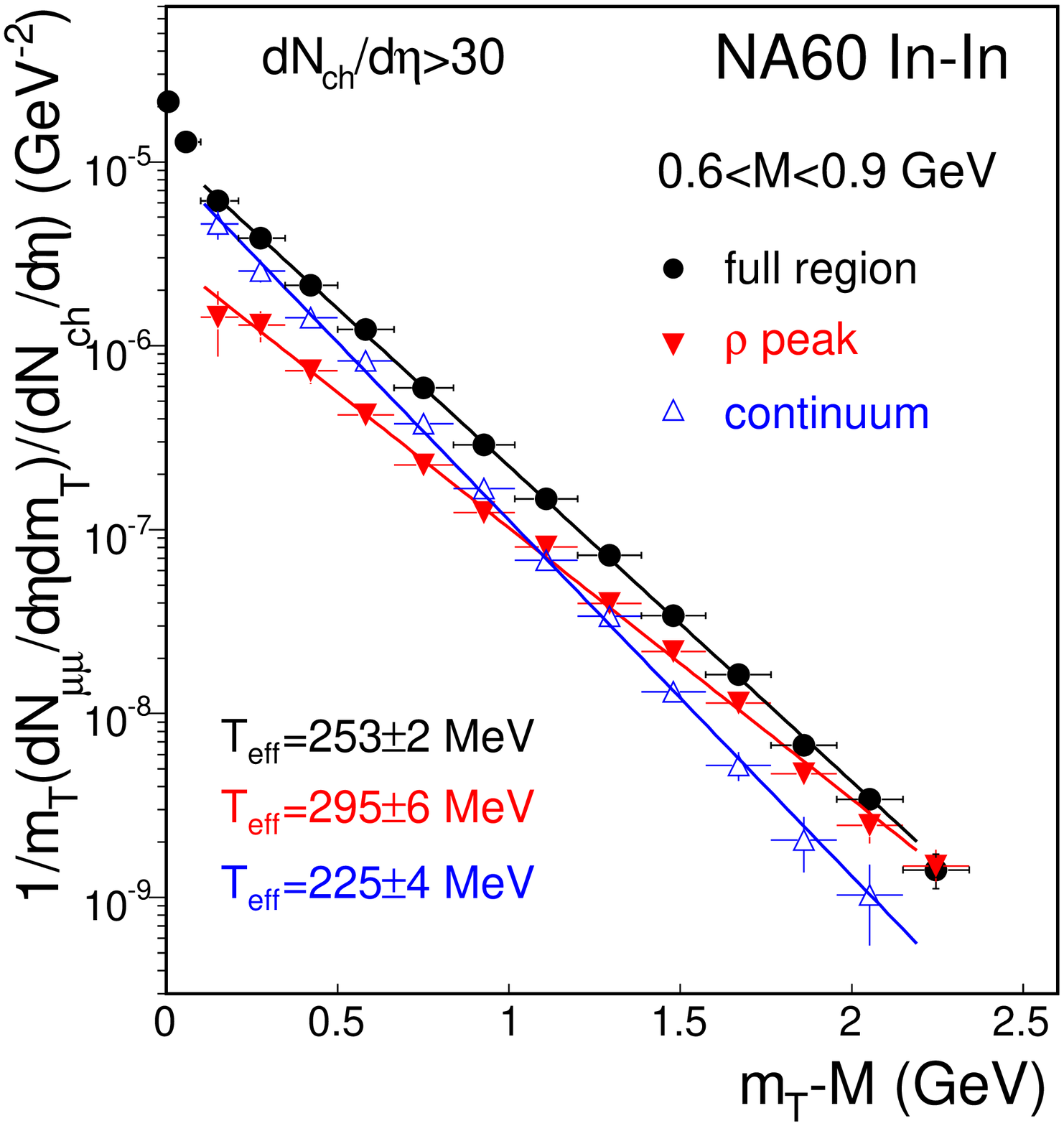}
\caption{Acceptance-corrected transverse mass spectra of the excess
dimuons for 4 mass windows and the $\phi$~\cite{Arnaldi:2007ru}
(left), and a decomposition into peak and continuum for the
$\rho$-like window (right, see text). The lines represent exponential
fits to the data in the range 0.4$<$p$_{T}$$<$1.8 GeV. The
normalization in absolute terms is independent of rapidity over the
region measured. For error discussion see~\cite{Arnaldi:2007ru}.}
   \label{fig4}
\end{center}
\end{figure*}

The two major variables characterizing dileptons are $M$ and $p_{T}$,
and the existence of two rather than one variable as in case of real
photons leads to much richer information. Beyond (minor) contributions
from the spectral function, $p_{T}$ encodes the key properties of the
expanding fireball, temperature and transverse (radial) flow. In
contrast to hadrons, however, which receive the full asymptotic flow
at the moment of decoupling, dileptons are continuously emitted during
the evolution, sensing the space-time development of temperature and
flow. This makes the dilepton $p_{T}$ spectra sensitive to the
emission region, providing a powerful diagnostic
tool~\cite{Kajantie:1986,Ruppert:2007cr}. Fig.~\ref{fig4} (left)
displays the centrality-integrated $m_{T}$ spectra, where $m_{T} =
(p_{T}^{2} + M^{2})^{1/2}$, for four mass windows; the $\phi$ is
included for comparison. The ordinate is normalized to $dN_{ch}/d\eta$
in absolute terms~\cite{Damjanovic:qm08,Damjanovic:hp08}. Apart from a
peculiar rise at low $m_{T}$ ($<$0.2 GeV) for the excess spectra (not
the $\phi$) which only disappears for very peripheral
collisions~\cite{Specht:2007ez,Arnaldi:2007ru}, all spectra are pure
exponentials, but with a mass-dependent slope. Fig.~\ref{fig4} (right)
shows a more detailed view into the $\rho$-like mass window, using the
same side-window method as described in connection with
Fig.~\ref{fig2} to determine the $p_{T}$ spectra separately for the
$\rho$ peak and the underlying continuum. All spectra are purely
exponential up to the cut-off at $p_{T}$=3 GeV, without any signs of
an upward bend characteristic for the onset of hard processes. Their
slopes are, however, quite different (see below).

\begin{figure}[]
\begin{center}
\includegraphics*[width=0.43\textwidth]{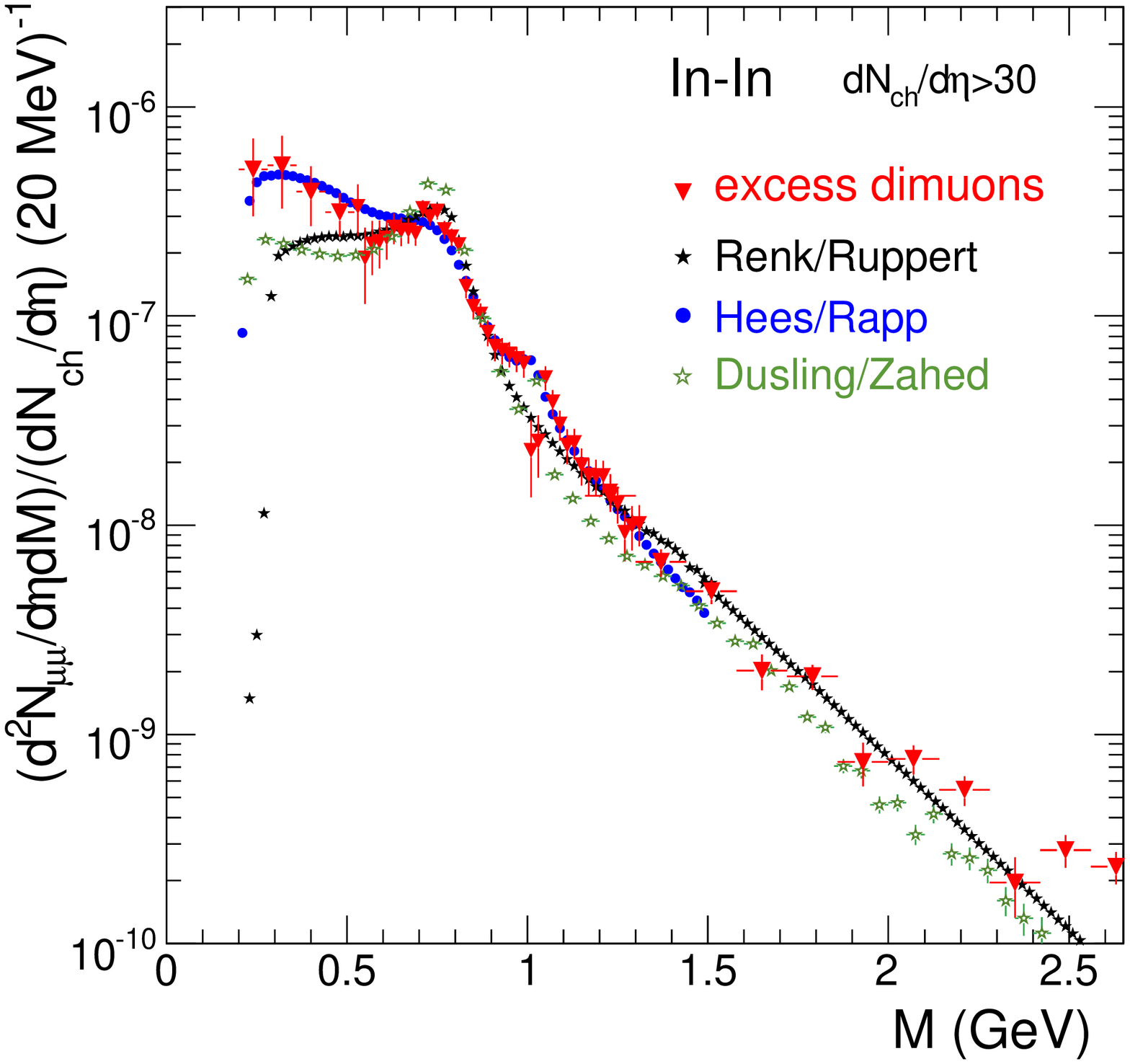}
\hspace*{0.7cm}
\includegraphics*[width=0.43\textwidth]{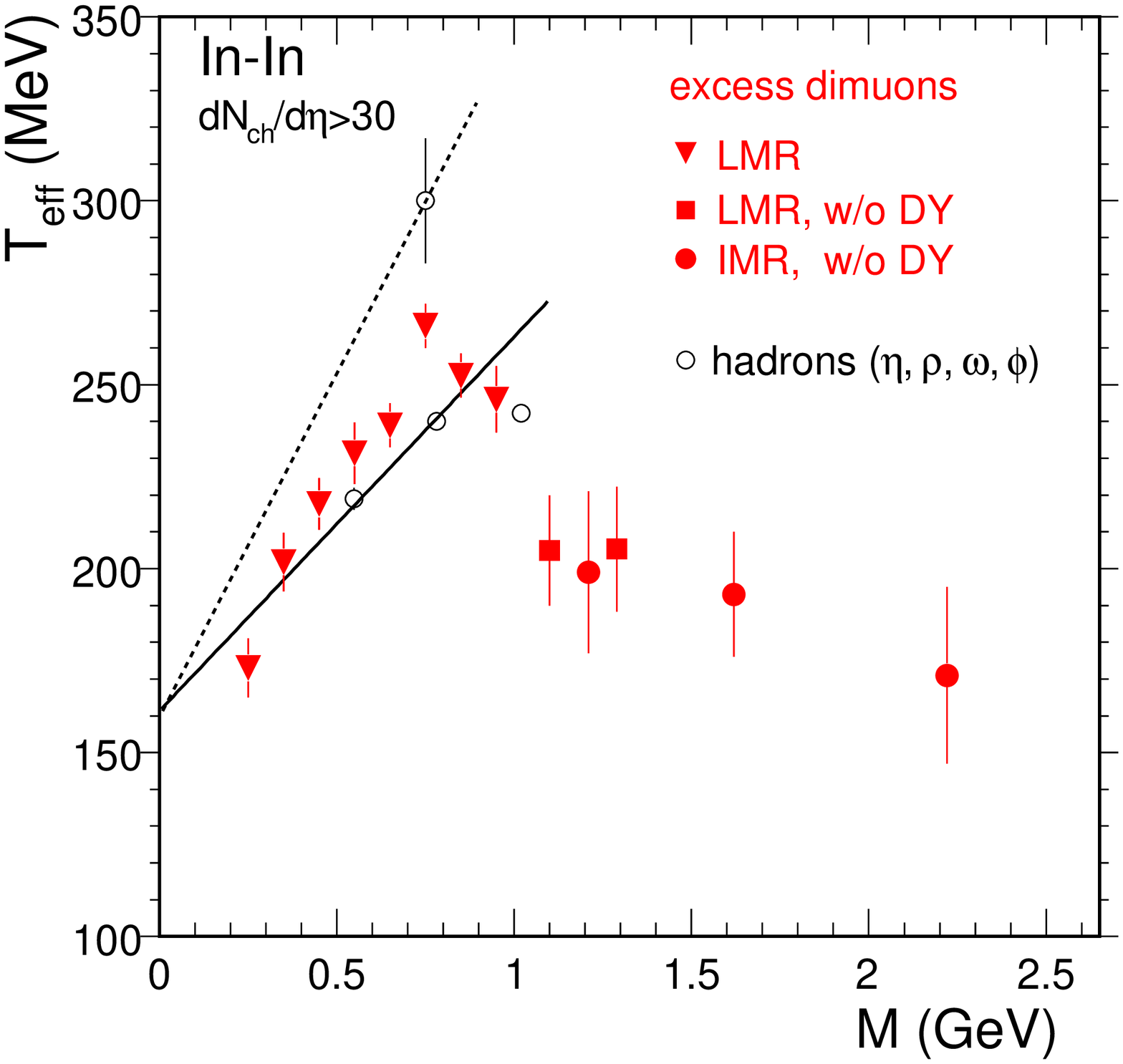}
\caption{Left: Acceptance-corrected mass spectra of the excess dimuons
for the combined LMR/IMR regions. The theoretical scenarios are
labeled according to the authors RR~\cite{Ruppert:2007cr},
HR~\cite{vanHees:2006ng}, and DZ~\cite{Dusling:2007rh}. In case
of~\cite{vanHees:2006ng}, the EoS-B$^{+}$ option is used. Right:
Inverse slope parameter $T_\mathrm{eff}$ vs. dimuon mass for the
combined LMR/IMR regions of the excess in comparison to
hadrons~\cite{Arnaldi:2007ru}. Open charm is subtracted in both parts
of the figure.}
  \label{fig5}
\end{center}
\end{figure}

The inverse slope parameters $T_\mathrm{eff}$ extracted from
exponential fits to the $m_{T}$ spectra (with a finer binning than in
Fig.~\ref{fig4}) are plotted in Fig.~\ref{fig5} (right) vs. dimuon
mass, unifying the data from the LMR and IMR regions. The hadron data
for $\eta$, $\omega$ and $\phi$ obtained as a by-product of the
cocktail subtraction are also included, as is the single value for the
$\rho$-peak from Fig.~\ref{fig4} (right). Interpreting the latter as
the freeze-out $\rho$ without in-medium effects, all four hadron
values together with preliminary $\pi^{-}$ data from NA60 can be
subjected to a simple blast-wave analysis~\cite{Damjanovic:qm08}. This
results in a reasonable set of freeze-out parameters of the fireball
evolution and suggests the following consistent interpretation of the
hadron and dimuon data together. Maximal flow is reached by the
$\rho$, due to its maximal coupling to pions, while all other hadrons
freeze out earlier. The $T_\mathrm{eff}$ values of the dilepton excess
rise nearly linearly with mass up to the $\rho$-pole position, but
stay always well below the $\rho$ line, exactly what would be expected
for {\it radial flow} of an {\it in-medium} {\it hadron-like} source
(here $\pi^{+}\pi^{-}$$\rightarrow$$\rho$) decaying continuously into
dileptons.

Beyond the pole region of the $\rho$, however, the $T_\mathrm{eff}$
values of the excess show a sudden decline by about 50
MeV. Extrapolating the lower-mass trend to beyond the $\rho$, such a
fast transition to a seeming low-flow situation is extremely hard to
reconcile with emission sources which continue to be of dominantly
hadronic origin in this region. A more natural explanation would then
be a transition to a dominantly early {\it partonic} source with
processes like $q\bar{q}\rightarrow \mu^{+}\mu^{-}$ for which flow has
not yet built up~\cite{Ruppert:2007cr}. While still
controversial~\cite{vanHees:2006ng}, this may well represent the first
direct evidence for thermal radiation of partonic origin, overcoming
parton-hadron duality for the {\it yield} description in the mass
domain.

The acceptance- and efficiency-corrected data can also be projected on
the mass axis. The p$_{T}$-differential results are discussed
in~\cite{Damjanovic:qm08,Damjanovic:hp08}. Fig.~\ref{fig5} (left)
shows p$_{T}$-integrated data for p$_{T}$$>$0.2 GeV (the lowest
p$_{T}$ bin has been excluded to avoid the very large errors in the
low-acceptance region~\cite{Damjanovic:qm08,Damjanovic:hp08}). The LMR
and IMR regions are again unified, using an independently assessed
absolute normalization~\cite{Ruben:panic08}. Recent theoretical
results from the three major groups working in the field are included
for comparison~\cite{Ruppert:2007cr,vanHees:2006ng,Dusling:2007rh}. A
strong rise towards low masses is seen in the data, reflecting the
Boltzmann factor, i.e. the Plank-like radiation associated with a very
broad, nearly flat spectral function. Only the Hees/Rapp
scenario~\cite{vanHees:2006ng} is able to describe this part
quantitatively, due to their particularly large contribution from
baryonic interactions to the low-mass tail of the $\rho$ spectral
function. The $\rho$ pole remains visible, due to the freeze-out
part. In-medium broadening of parts of the $\omega$ and $\phi$ are
contained in~\cite{vanHees:2006ng}, but not
in~\cite{Ruppert:2007cr,Dusling:2007rh}, accounting for the broad bump
seen in the region of the $\phi$. Caution should, however, be taken as
to the nearly quantitative agreement with the data, since some details
of the resolution function of the NA60 apparatus are still under
investigation.

\vspace*{0.7cm}

\noindent{\bf Real photons and low-mass dileptons}
\vspace*{0.2cm}

Processes contributing to dilepton production at very low masses
include $\pi\pi$ annihilation (extending to below 2M$_{\pi}$ because
of the softening of the pion dispersion relation), $\pi$-baryon
processes, the a$_{1}$-Dalitz decay, parton annihilation, internal
conversion of photons from processes creating also real photons
(e.g. Compton-like graphs or
\begin{figure}[h!]
\begin{minipage}[l][8.0cm][t]{9cm}
\includegraphics*[width=1.1\textwidth,bb = 1 3 560 454]{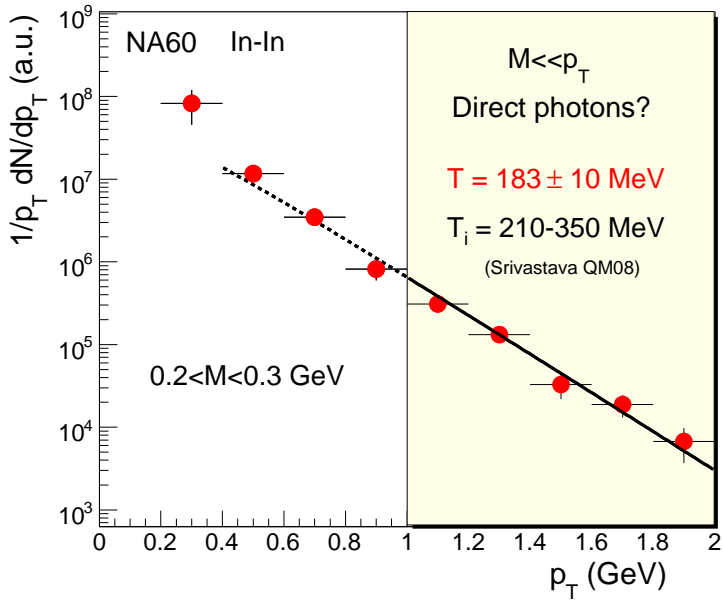}
\end{minipage} \qquad\qquad
\begin{minipage}[l][8.0cm][t]{7cm}
\vspace*{3.2cm}
\caption{Acceptance-corrected p$_{T}$ spectrum of the excess for the
lowest mass bin 0.2$<$M$<$0.3 GeV. The solid line corresponds to an
exponential fit (as in Fig.~\ref{fig4}) for p$_{T}$$>$1 GeV, while the
dashed line is just an extrapolation down to p$_{T}$$<$1 GeV.}
   \label{fig6}
\end{minipage}
\end{figure}
bremsstrahlung), and
others~\cite{Ruppert:2007cr,vanHees:2006ng,Dusling:2007rh}, and the
superposition of all these sources accounts for the total measured
yield as shown in Figs.~\ref{fig1} and~\ref{fig5}. PHENIX at RHIC has
recently~\cite{Phenix:2008fqa} analysed dielectron data in the mass
window 0.2$<$M$<$0.3 GeV for p$_{T}$$>$1 GeV under the extreme
assumption, that the total yield for that selection would solely be
due to internal conversion of real photons; the transformation back to
real photons was done with the prescription
in~\cite{KrollWada1955}. The analogous situation for NA60 is
illustrated in Fig.~\ref{fig6}, but with the complete
p$_{T}$-spectrum; it corresponds to the lowest point in
Fig.~\ref{fig5} (right). The fit value for the cut $p_{T}$$>$1 GeV is
$T$=183$\pm$10 MeV. However, this value of $T$ corresponds to the
effective temperature $T_\mathrm{eff}$ as used in Fig.~\ref{fig5},
affected by radial flow (ignored by PHENIX). If corrected for that,
using the intercept of the linear rise with the ordinate, it is
reduced to about 150 MeV. Such a low value hardly makes sense in the
light of initial temperatures of 210-350 MeV expected for real
photons~\cite{Srivastava:qm08}, but is more indicative for the average
temperature of the hadronic phase alone. In Fig.~\ref{fig5}, the
lowest point does not look exceptional at all, but just appears as
part of the general systematic rise of $T_\mathrm{eff}$ vs. mass. The
lack of any convincing tag on internal conversion and the embeddment
of the lowest mass bin in the general systematics rule out the real
photon hypothesis for this bin, and this will hardly change at higher
masses.

\vspace*{0.4cm}
\noindent{\bf Conclusions}
\vspace*{0.2cm}

Excess dileptons have been observed over the complete mass region
0.2$<$M$<$2.6 GeV. Their p$_{T}$ spectra offer a clear view of the
origin of the different sources. All observations, including the
finding of zero polarization, are consistent with a global
interpretation of the excess as thermal radiation. There is no way to
unambiguously isolate real photons via internal conversion of low-mass
dileptons.

\end{document}